\documentclass[journal,twoside]{IEEEtran}

\usepackage{color}
\usepackage{bbm}
\usepackage{url}
\usepackage[cmex10]{amsmath}
\usepackage{amsfonts,amssymb}
\usepackage{dsfont}
\usepackage{graphicx}
\usepackage[hang]{subfigure}
\usepackage{citesort}
\usepackage{mathtools}
\usepackage{bbm}
\usepackage{fixmath}
\usepackage{tcolorbox}
\usepackage{balance}
\usepackage{manfnt}

\usepackage{ntheorem}
\theoremstyle{plain}
\theoremheaderfont{\itshape}\theorembodyfont{\itshape}
\theoremseparator{.}
\newtheorem{theorem}{Theorem}

\theoremheaderfont{\itshape}\theorembodyfont{\upshape}
\theoremseparator{.}
\newtheorem{definition}{Definition}
\newtheorem{remark}{Remark}

\renewcommand{\vec}[1]{\boldsymbol{#1}}

\newcommand{\N}{\mathbb{N}}
\newcommand{\R}{\mathbb{R}}

\newcommand{\sA}{\mathcal{A}}

\newcommand{\sC}{\mathcal{C}}

\newcommand{\sI}{\mathcal{I}}

\newcommand{\sM}{\mathcal{M}}
\newcommand{\sN}{\mathcal{N}}
\newcommand{\sP}{\mathcal{P}}

\newcommand{\sS}{\mathcal{S}}
\newcommand{\sT}{\mathcal{T}}
\newcommand{\sU}{\mathcal{U}}

\newcommand{\sX}{\mathcal{X}}
\newcommand{\sY}{\mathcal{Y}}

\newcommand{\fW}{\mathfrak{W}}

\newcommand{\CH}{\mathcal{CH}}

\newcommand{\Mdenial}{\mathcal{M}_{\text{DoS}}}
\newcommand{\Mdenialfull}{\overline{\mathcal{M}}_{\text{DoS}}}

\newcommand{\Msyms}{\sM_{\text{sym}}^<(P,\Lambda)}
\newcommand{\Msymse}{\sM_{\text{sym}}^\leq(P,\Lambda)}
\newcommand{\Msymge}{\sM_{\text{sym}}^\geq(P,\Lambda)}

\newcommand{\BSS}{\mathfrak{B}}

\newcommand{\Mdos}{\mathcal{M}_{\text{DoS}}}
\newcommand{\Mdosfull}{\overline{\mathcal{M}}_{\text{DoS}}}
\newcommand{\Tdos}{\mathcal{T}_{\text{DoS}}}
\newcommand{\Tnodos}{\mathcal{T}_{\text{noDoS}}}
\newcommand{\Pos}{\mathcal{P}_{\text{os}}}
\newcommand{\CHxys}{\CH(\sX,\sS;\sY)}
\newcommand{\CHxs}{\CH(\sX;\sS)}

\newcommand{\addspace}{\vspace*{0.25\baselineskip}}

\begin{document}

\title{On the Need of Neuromorphic Twins \\ to Detect Denial-of-Service Attacks \\ on Communication Networks}
\author{Holger Boche,~\IEEEmembership{Fellow,~IEEE}, 
	Rafael F. Schaefer,~\IEEEmembership{Senior Member,~IEEE}, \\ 
	H. Vincent Poor,~\IEEEmembership{Life Fellow,~IEEE}, and
	Frank H. P. Fitzek,~\IEEEmembership{Senior Member,~IEEE}
	\thanks{This work of H. Boche was supported in part by the German Federal Ministry of Education and Research (BMBF) within the national initiative for ``\emph{Post Shannon Communication (NewCom)}'' under Grant 16KIS1003K and within the national initiative on 6G Communication Systems through the research hub 6G-life under Grant 16KISK002, as well as in part by the German Research Foundation (DFG) as part of Germany’s Excellence Strategy -- EXC-2092 - Project ID 390781972. This work of R. F. Schaefer was supported in part by the BMBF within NewCom under Grant 16KIS1004 and in part by the DFG under Grant SCHA 1944/6-1. This work of H. V. Poor was supported by the U.S. National Science Foundation under Grant CCF-1908308. This work of F. H. P. Fitzek was funded by the DFG as part of Germany’s Excellence Strategy -- EXC 2050/1 - Project ID 390696704 - Cluster of Excellence ``Centre for Tactile Internet with Human-in-the-Loop'' (CeTI) of Technische Universit\"at Dresden. This article was presented in part at the IEEE International Conference on Acoustics, Speech and Signal Processing (ICASSP), Toronto, ON, Canada, June 2021~\cite{BocheSchaeferPoor-2021-ICASSP-RealNumberSignalProcessingDoS}.
	}
	\thanks{Holger Boche is with the Institute of Theoretical Information Technology, Technische Universit\"at M\"unchen, 80290 Munich, Germany, the BMBF Research Hub 6G-life, and the Excellence Cluster Cyber Security in the Age of Large-Scale Adversaries (CASA), Ruhr University Bochum, 44801 Bochum, Germany (email: boche@tum.de).}
	\thanks{Rafael F. Schaefer is with the Chair of Information Theory and Machine Learning, Technische Universit\"at Dresden, 01062 Dresden, Germany (email: rafael.schaefer@tu-dresden.de).}
	\thanks{H. Vincent Poor is with the Department of Electrical and Computer Engineering, Princeton University, Princeton, NJ 08544, USA (email: poor@princeton.edu).}
	\thanks{Frank H. P. Fitzek is with the Deutsche Telekom Chair of Communication Networks, Technische Universit\"at Dresden, 01187 Dresden, Germany, and the Cluster of Excellence ``Centre for Tactile Internet with Human-in-the-Loop'' (CeTI) (email: frank.fitzek@tu-dresden.de).}         
}

\IEEEoverridecommandlockouts
\maketitle

\begin{abstract}
    As we are more and more dependent on the communication technologies, resilience against any attacks on communication networks is important to guarantee the digital sovereignty of our society. New developments of communication networks tackle the problem of resilience by in-network computing approaches for higher protocol layers, while the physical layer remains an open problem. This is particularly true for wireless communication systems which are inherently vulnerable to adversarial attacks due to the open nature of the wireless medium. In \emph{denial-of-service (DoS)} attacks, an active adversary is able to completely disrupt the communication and it has been shown that Turing machines are incapable of detecting such attacks. As Turing machines provide the fundamental limits of digital information processing and therewith of digital twins, this implies that even the most powerful digital twins that preserve all information of the physical network error-free are not capable of detecting such attacks. This stimulates the question of how powerful the information processing hardware must be to enable the detection of DoS attacks. Therefore, in the paper the need of \emph{neuromorphic twins} is advocated and by the use of \emph{Blum-Shub-Smale machines} a first implementation that enables the detection of DoS attacks is shown. This result holds for both cases of with and without constraints on the input and jamming sequences of the adversary.  

\end{abstract}


\section{Introduction}
\label{sec:intro}

The fifth generation (5G) of mobile communication systems is boosting the digital transformation of our society as it expanded the use cases from consumer-centric to machine-centric. New applications such as driverless mobility, health care, constructions, agriculture, and many more opened new markets for network operators and manufacturers. With 6G the use cases even expanded from not only machine-to-machine but also machine-to-human often referred to as the Tactile Internet \cite{CeTIBook-2020q} or the Metaverse.
The new communication technologies enable our society to cope with the biggest ubiquitous problems such as the climate crisis, energy crisis, or local conflicts.

But as we are more and more dependent on the communication technologies, resilience against any failure or intended attackers of communication networks becomes even more important to guarantee the digital sovereignty of our society.
Resilience as a core part of trustworthiness has been identified as a key challenge for 6G and must be understood and addressed at an unprecedented level \cite{Cherkaoui-2021-6G}. With the aforementioned critical applications comes the need to address the trustworthiness of the system and its services. The theory for the formalization of trustworthiness must comprise different fields \cite{FettweisBoche-2022-ProcACM-6GTrust} and many central questions and issues are open to date. Most attacks are performed through the communication network by exploiting inherent weaknesses of the communication protocol on higher layers.

To face these challenges, there is the recent trend towards shifting functionalities from the physical layer to higher layers by enabling software-focused solutions. Software solutions have the advantage of short update cycles, adaption to new situations, and lower costs with respect to CAPEX and OPEX. The aim is to create an infrastructure that is capable of interconnecting highly heterogeneous networks to support several different verticals.

Promising approaches include the concepts of software-defined networking (SDN) \cite{Nunes-2014-COMST-SurveySDN} and network function virtualization (NFV) \cite{Mijumbi-2016-COMST-NFV}. The aim of such network virtualization is to provide software-based solutions for functions, protocols, and operations such that they run on general purpose hardware and do not require specialized hardware anymore. The placement of the aforementioned network functions can be realised by the SDN-controller and is often optimized to minimize the communication latency. This had laid ground for novel paradigms such as cloud and edge/fog computing, unique and reconfigurable SDN-NFV architectures and end-to-end network slicing \cite{Richart-2016-TNSM-ResourceSlicingVirtualWirelessNetworks,Li-2017-IC-NetworkSlicing5G}.
With this comes the need of addressing the trustworthiness and security threads due to network softwarization and the problem of resilience is tackled by in-network computing appraoches for higher protocol layer \cite{Ahmad-2015-COMST-SecuritySDN,Dargahi-2017-COMST-SurveySecuritySDNDataPlanes,Farris-2019-COMST-SurveyEmergingSDNNFVSecurity,ScottHaywardNatarajanSezer-2016-COMST-SurveySecuritySDN,AbdouVanOorschotWan-2018-COMST-AnalysisControlPlaneSecuritySDN,Pattaranantakul-2018-COMST-NFVSecuritySurvey}. Initial studies on the algorithmic verification of trustworthiness based on digital hardware platforms are given in \cite{BocheSchaeferPoorFettweis-2022-ICC-Trustworthiness}.

The physical layer remains an open problem and, accordingly, this paper studies attacks that are performed directly at the physical layer with the aim of disrupting the physical transmission itself. Such attacks can target a specific single user within the system, but also the overall system itself. Reliable communication between legitimate users is the indispensable basis for any information processing. And particularly wireless communication systems are inherently vulnerable to adversarial attacks due to the open nature of the wireless medium and the high number of co-existing wireless systems which easily disturb each other. Moreover, malevolent jammers might even jam and harm the legitimate communication intentionally. 

\begin{figure}
	\centering
	\scalebox{1}{\includegraphics{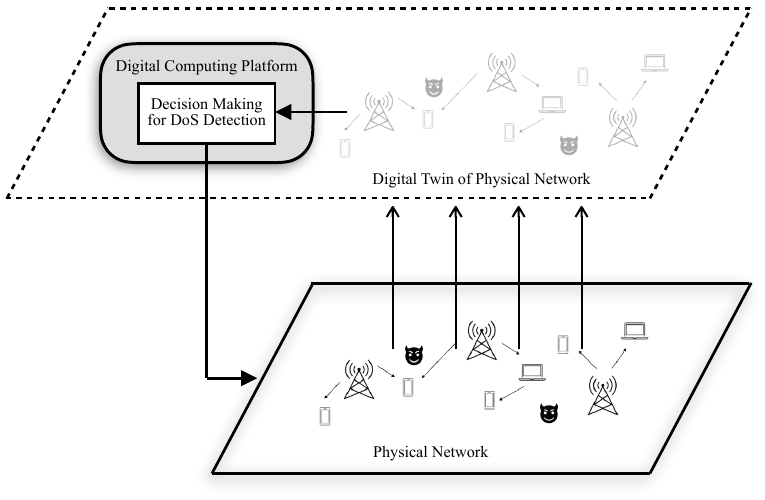}}
	\caption{A digital twin creates a digital description of the physical network which is then used by appropriate digital computing platform for the detection of DoS attacks. Here, an optimal digital twin provides a complete characterization of the physical system without any information loss.}
	\label{fig:digitaltwin1}
\end{figure}

In \emph{denial-of-service (DoS)} attacks, an active adversary is able to completely disrupt the communication and it is of utmost importance to detect such attacks. In the end, the detection must be done \emph{algorithmically} on the deployed information processing hardware. 

Within the framework of \emph{digital twins}, a suitable description of the physical network is created which is then used by the corresponding digital computing platform for the detection task, cf. Fig.~\ref{fig:digitaltwin1}. In the best case, this digital twin provides the complete description of the physical network, i.e., all parameters and components of the physical system are completely characterized by a digital description. To this end, all parameters and components of the physical system are described by programs that provide suitable digital objects for these for every given approximation error. This means that an optimal digital twin provides a description of the analog physical system without any information loss. In the following, we will assume such optimal digital twins.

To date, practical information processing hardware platforms are digital and \emph{Turing machines} \cite{Turing-1936-ComputableNumbersEntscheidungsproblem,Turing-1937-ComputableNumbersEntscheidungsproblemCorrection,Weihrauch-2000-ComputableAnalysis} provide fundamental performance limits for today’s digital information processing and therewith immediately also for digital twins. Accordingly, Turing machines are therefore the ideal framework to study whether or not it is in principle possible to detect DoS attacks with the help of digital twins. Thus, a suitable Turing machine needs to decide whether or not an adversary is able to perform a DoS attack for a given communication scenario. A negative answer has been obtained in \cite{BocheSchaeferPoor-2020-TSP-DoS}, where it has been shown that there is no Turing machine that can algorithmically detect a DoS attack; even for such optimal descriptions by optimal digital twins. This shows that even with the most powerful digital twin and the most powerful digital computing platform (given by Turing machines), the detection of DoS attacks is impossible. 

Turing machines can only operate with rational numbers and computable real numbers are then those that can be effectively approximated by computable sequences of rational numbers. 
The non-detectability of DoS attacks stimulate the questions of whether this comes from the fact that Turing machines approximate computable real numbers with the help of rational numbers and how powerful a twin and the corresponding information processing hardware platform must be to enable a successful detection of DoS attacks. 

In this paper, we advocate the need of \emph{neuromorphic twins} that are based on neuromorphic computing hardware platforms. 
Obviously, it is not clear if these can be realized with the hardware technology that used to date in communication systems, since it would require the processing and storage of arbitrary real numbers. 

Over the last years, new electronic hardware platforms for neuromorphic computing have been developed and particularly the industry (such as IBM, Intel, or Samsung) made some significant progress in the development of powerful neuromorphic processors. A suitable computing model for this had been proposed in 1989 by Blum, Shub, and Smale \cite{BlumShubSmale-1989-AMS-TheoryComputation}. 
They proposed a general computing model over an arbitrary ring or field. Note that if this is chosen to be $\mathbb{Z}_2=\{\{0,1\},+,\cdot\}$, then these \emph{Blum-Shub-Smale (BSS) machines} recover the theory of Turing machines. In addition, there are very interesting connections to the theory of semialgebraic sets.

BSS machines are further relevant for the field of biocomputing. In \cite{Grozinger-2019-Biocomputing} it has been conjectured that BSS machines provide the basis for biocomputing. Biocomputing covers all different scales of computing in living organisms ranging from computing in cells to computing in the brain. 

In this paper, we consider such BSS machines and study whether or not these are capable to detect DoS attacks. We prove that such attacks are indeed detectable by BSS machines showing that an implementation based on neuromorphic twins enables the algorithmic detection of DoS attacks. This result is shown to hold for both cases of with and without constraints on the input and jamming sequences of the adversary. For this purpose, it is further shown that the set of all channels for which a DoS attack is possible, is a semialgebraic set and can therefore be described by finitely many polynomial equations and inequalities.

\subsection*{Notation}

Discrete random variables are denoted by capital letters and their realizations and ranges by lower case and calligraphic letters, respectively; $\N$ and $\R$ are the sets of non-negative integers and real numbers; $\sP(\sX)$ denotes the set of all probability distributions on $\sX$ and $\CH(\sX;\sY)$ denotes the set of all stochastic matrices (channels) $\sX\rightarrow\sP(\sY)$.

\section{System Model and Problem Formulation}
\label{sec:model}

In this section, we introduce the communication system model and formulate the main problem.

\subsection{Communication System Model}
\label{sec:model_comsystem}

\begin{figure}
	\centering
	\scalebox{1}{\includegraphics{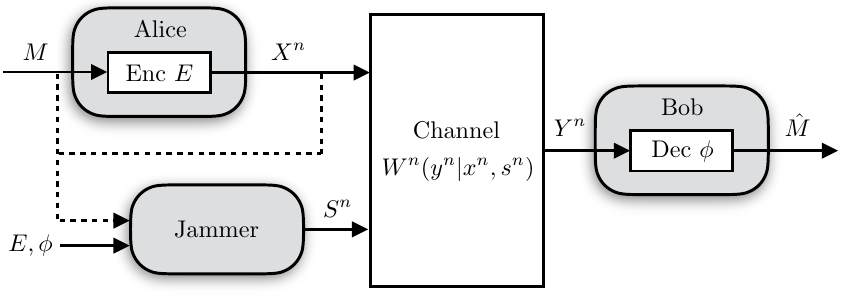}}
	\caption{Communication system with an active adversary as in \cite{BocheSchaeferPoor-2020-TSP-DoS}. The Jammer tries to disrupt the communication between Alice and Bob by sending an own jamming sequence. A \emph{Jammer with partial knowledge} knows the encoder $E$ and decoder $\phi$ and can choose its jamming strategy accordingly. A \emph{Jammer with full knowledge} is further aware of the actual message $M$ (or the codeword $X^n=X^n(M)$) and can adapt its jamming sequence to each particular message accordingly.}
	\label{fig:model}
\end{figure}

The communication system of interest is shown in Fig.~\ref{fig:model}. Here, a transmitter Alice wants to transmit a message reliably to a receiver Bob, while an active adversary (or Jammer) tries to disrupt the transmission by sending an own jamming sequence. To this end, the Jammer can choose this jamming sequence based on its knowledge about the legitimate communication. A \emph{Jammer with partial knowledge} is aware of the overall system design, i.e., the encoder and decoder used by Alice and Bob, respectively, while a \emph{Jammer with full knowledge} further knows the actual message that is transmitted. We also refer to \cite{BocheSchaeferPoor-2020-TSP-DoS} for a detailed introduction and discussion on these two cases. 

As Alice and Bob have no prior knowledge about the jamming strategy of the Jammer and how the channel will be disturbed, they have to be prepared for the worst: A channel that may vary in an arbitrary and unknown manner from channel use to channel use. The concept of \emph{arbitrarily varying channels (AVCs)} \cite{Blackwell-1960-CapacitiesAVCRandomCoding,Ahlswede-1978-EliminationCorrelationAVC,CsiszarNarayan-1988-AVCRevisited} is a suitable model to capture the effects of such unknown varying channel conditions. 

To this end, let $\sX$ and $\sY$ be finite input and output alphabets and $\sS$ a finite state (jamming) alphabet. The channel from Alice to Bob is then given by a stochastic matrix $W:\sX\times\sS\rightarrow\sP(\sY)$ which we interchangeably write as $W\in\CHxys$. For a fixed jamming sequence $s^n\in\sS^n$ of length $n$, the discrete memoryless channel is $W^n(y^n|x^n,s^n)\coloneqq\prod_{i=1}^{n}W(y_i|x_i,s_i)$ for all input and output sequences $x^n\in\sX^n$ and $y^n\in\sY^n$.

\begin{definition}
	\label{def:avc}
	The \emph{arbitrarily varying channel (AVC)} $\fW$ is given by
	\begin{equation*}
		\fW = \big\{W(\cdot|\cdot,s)\big\}_{s\in\sS}.
	\end{equation*}
\end{definition}

For the characterization of the capacity of an AVC, we need the concept of \emph{symmetrizability} \cite{CsiszarNarayan88AVCRevisited}.

\begin{definition}
	\label{def:sym}
	An AVC is called \emph{symmetrizable} if there exists a stochastic matrix $U:\sX\rightarrow\sP(\sS)$ such that
	\begin{equation*}
		\sum_{s\in\sS}W(y|x,s)U(s|\hat{x}) = \sum_{s\in\sS}W(y|\hat{x},s)U(s|x)
	\end{equation*}
	holds for all $x,\hat{x}\in\sX$ and $y\in\sY$.
\end{definition}

Intuitively, this captures the ability of the Jammer to ``emulate'' valid channel inputs making it impossible for the receiver to distinguish whether the channel output comes from a valid codeword sent by Alice or the jamming input of the Jammer.

\subsubsection{Jammer with Partial Knowledge}
\label{sec:model_comsystem_partial}

This accounts for all those jammers that know the encoding and decoding functions of the legitimate users, but not the transmitted message. This is captured very well by an AVC under the average error criterion, cf. also \cite{BocheSchaeferPoor-2020-TSP-DoS} for further details. In this case, the concept of symmetrizability suffices to completely characterize the capacity.

\begin{theorem}[{\cite{Ahlswede78EliminationCorrelationAVC,CsiszarNarayan88AVCRevisited}}]
	\label{the:avc}
	The capacity $C(\fW)$ of an AVC $\fW$ under the average error criterion is
	\begin{align*}
		C(\fW) = \begin{cases}
			\min_{q\in\sP(\sS)}C(W_q) &\text{if }\fW\text{ is non-symmetrizable} \\	
			0 &\text{if }\fW\text{ is symmetrizable} 
		\end{cases}
	\end{align*}
	with $C(W_q)$ the regular Shannon capacity of the averaged channel $W_q(y|x)=\sum_{s\in\sS}W(y|x,s)q(s)$, $q\in\sP(\sS)$.
\end{theorem}

In this paper, we are particularly interested in DoS attacks for which the Jammer is able to completely disrupt the communication. We denote the set of all these channels for which a DoS attack is possible by $\Mdos$. Then Theorem~\ref{the:avc} immediately reveals that such attacks are exactly possible whenever the channel is symmetrizable so that 
\begin{equation*}
	\Mdos=\big\{\fW:\fW\text{ is symmetrizable}\big\}
\end{equation*}
and $\Mdos^c=\{\fW:\fW\text{ is non-symmetrizable}\}$. Thus, it immediately allows to analytically decide whether or not DoS attacks are possible.

\subsubsection{Jammer with Full Knowledge}
\label{sec:model_comsystem_full}

This accounts for all those jammers that know the encoding and decoding functions of the legitimate users and that are further aware of the actual transmitted message. This can be modeled by an AVC under the maximum error criterion, cf. also \cite{BocheSchaeferPoor-2020-TSP-DoS} for further details. Although the capacity is not known in this case, a condition has been established (similar to the symmetrizability condition in the average error case) that allows to characterize exactly when the capacity is zero and therewith when a DoS attack is possible.

\begin{theorem}[{\cite{Ahlswede78EliminationCorrelationAVC}}]
	For an AVC under the maximum error criterion, we have for the capacity $C_{\text{max}}(\fW)>0$ if and only if there exists $x,\hat{x}\in\sX$ with
	\begin{equation}
		\label{eq:avcmax}
		\sI(\fW,x)\cap \sI(\fW,\hat{x}) = \emptyset
	\end{equation}
	where
	\begin{equation}
	\begin{split}
	    \label{eq:avcmax2}
		\sI(\fW,x) = \Big\{p\in\sP(\sY): &\exists q\in\sP(\sS) \\
		&\text{ s.t. } p(y)=\sum_{s\in\sS}W(y|x,s)q(s)\Big\}.
	\end{split}
	\end{equation}
\end{theorem}

Similarly as above, we denote the set of all channels for which a DoS attack is possible by $\Mdosfull$ and we see that $\Mdosfull$ can be characterized with the help of \eqref{eq:avcmax} and \eqref{eq:avcmax2}. Thus, also in this case, we can analytically immediately decide whether or not a DoS attack is possible.

\subsection{Problem Formulation}
\label{sec:model_problem}

\begin{figure}
	\centering
	\scalebox{1}{\includegraphics{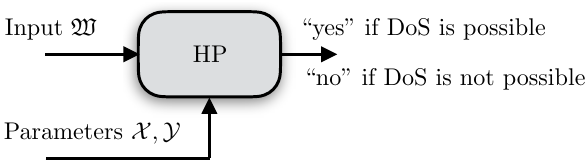}}
	\caption{Information processing hardware platform for the algorithmic detection of DoS attacks. It gets the parameters $\sX$, $\sY$, and the AVC $\fW$ as inputs and then computes whether or not a DoS attack is possible. At this point, the hardware platform is not further specified and can be either digital, i.e., given by a Turing machine, or, for example, neuromorphic which would then lead to a BSS machine.}
	\label{fig:hp}
\end{figure}

The previous discussion and results show that it can \emph{analytically} be decided whether or not DoS attacks are possible. The question is now whether or not such DoS attacks can also be detected \emph{algorithmically}, i.e., on a suitable information processing hardware platform as visualized in Fig.~\ref{fig:hp} and with the help of digital twins as initially motivated in Fig.~\ref{fig:digitaltwin1}. This leads to the following question. 
\addspace

\begin{tcolorbox}[colback=white,boxrule=0.125ex]
	{\bf Question:} Is there an information processing hardware platform that takes the channel and its parameters as inputs and then algorithmically decides whether or not a DoS attack is possible?
\end{tcolorbox}

We emphasize that this question has been posed in a very general form in the sense that the underlying hardware platform has not been specified so far. To date, practical information processing hardware platforms are digital including digital signal processing (DSP), field gate programmable array (FPGA) platforms, and even current supercomputers. As Turing machines provide the fundamental performance limits for today's digital information processing hardware platforms, they are therefore the ideal concept to study whether or not DoS attacks can be detected on digital hardware platforms in general.

The algorithmic detection of DoS attacks on Turing machines has been studied in detail in \cite{BocheSchaeferPoor-2020-TSP-DoS}. Unfortunately, the following negative answer has been obtained.

\begin{theorem}[{\cite{BocheSchaeferPoor-2020-TSP-DoS}}]
	\label{the:turing}
	Let $\sX$, $\sY$, and $\sS$ be arbitrary finite alphabets. Then there is no Turing machine that can detect a DoS attack. This result holds for both cases of jammers with partial and full knowledge.
\end{theorem}

\begin{figure}
	\centering
	\scalebox{1}{\includegraphics{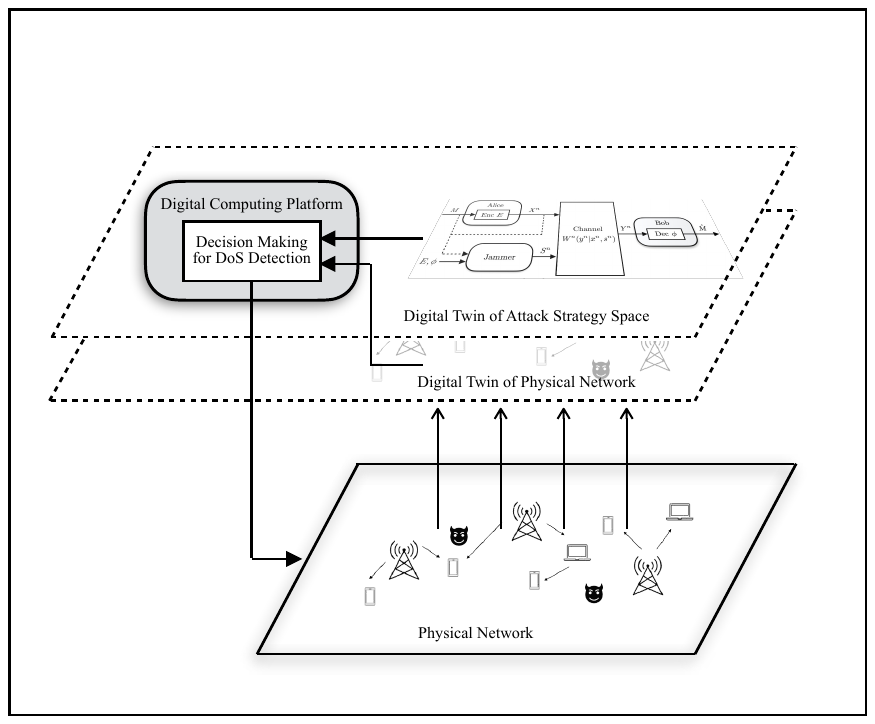}}
	\caption{In addition to the digital twin of the physical network, a second digital twin is employed that provides a digital description of the attack strategy space in addition to the description of the physical network. Both are then used for the detection of DoS attacks.}
	\label{fig:digitaltwin2}
\end{figure}

This shows that the detection of DoS attacks remains impossible with the hardware technology that is used to date in communication systems. In addition, this also shows the fundamental limitations of digital twins for such tasks. Note that this includes even the more general case where a second digital twin is employed that further provides a suitable characterization of the attacker's strategy space as visualized in Fig.~\ref{fig:digitaltwin2}.

This and the industry's progress in the development of neuromorphic processors stimulate the question of how powerful the information processing hardware and corresponding twins must be to enable such a DoS detection. To account for this, we consider in the following a more general computation model based on BSS machines that allows the processing and storage of arbitrary real numbers. We will show that such BSS machines and therewith corresponding \emph{neuromorphic twins} will always enable the detection of DoS attacks. A detailed description of such neuromorphic twins will be given in Section \ref{sec:dos}, but the crucial part is the following: We map the complete problem of DoS detection from the physical system into neuromorphic information and neuromorphic computing hardware. This includes the description of the physical network into corresponding neuromorphic ones, i.e., a neuromorphic twin and neuromorphic computing.

\section{BSS Machines and Semialgebraic Sets}
\label{sec:bss}

Here we introduce and discuss BSS machines and semialgebraic sets. These concepts will be needed subsequently.

\subsection{Blum-Shub-Smale Machines}
\label{sec:bss_bss}

The BSS machine has been introduced by Blum, Shub, and Smale in 1989 \cite{BlumShubSmale-1989-AMS-TheoryComputation}. This concept is a suitable candidate of a general computing model that reflects the implicit assumptions on the computability of capacity expressions. The BSS machine can store arbitrary real numbers, can compute all field operations on $\R$, i.e., ``$+$'' and ``$\cdot$'', and can compare real numbers according to the relations ``$<$'', ``$>$'', and ``$=$''. 

At the top level, a BSS machine is similar to a Turing machine in the sense that it operates on an infinite strip of tape according to a so-called program. This is a finite directed graph with five types of nodes associated with different operations: input node, computation node, branch node, shift node, and output node. For a detailed introduction and description of BSS machines and programs running on BSS machines, we refer the reader to \cite{Blum-2004-AMS-ComputingOverReals,BlumCuckerShubSmale-1998-Springer-ComplexityRealComputation} and references therein. 

\begin{definition}
	\label{def:computable}
	\emph{BSS-computable} functions are input-output maps $\Phi$ of the BSS machine $\BSS$, i.e., for every input $\vec{x}$, the output $\Phi_\BSS(\vec{x})$ is defined if the ouput is reachable by the program of the BSS machine $\BSS$.
\end{definition}

In the sequel, we will further need the concepts of BSS-decidability and BSS-semidecidability which are defined next.

\begin{definition}
	\label{def:decidable}
	A set $\sA\subset\R^N$ is \emph{BSS-decidable} if there is a BSS machine $\mathfrak{BSS}_\sA$ such that for all $\vec{x}\in\R^N$ we have
	$\mathfrak{BSS}_\sA(\vec{x}) = \chi_\sA(\vec{x})$,
	i.e., the characteristic function $\chi_\sA$ of the set $\sA$ is BSS-computable. 
\end{definition}

\begin{definition}
	\label{def:semidecidable}
	A set $\sA\subset\R^N$ is \emph{BSS-semidecidable} if there is a BSS machine $\mathfrak{BSS}_\sA$ with only one output (halting) state and that stops for input $\vec{x}\in\R^N$ if and only if $\vec{x}\in\sA$. 
\end{definition}

Some remarks are in order:
\begin{enumerate}
	\item A set $\sA$ is BSS-semidecidable if and only if there is a BSS machine that has only one halting state and that accepts only $\vec{x}\in\sA$.
	\item A set $\sA$ is BSS-decidable if and only if $\sA$ and $\sA^c$ are both BSS-semidecidable.
\end{enumerate}

\subsection{Semialgebraic Sets}
\label{sec:bss_semi}

We further need the concept of semialgebraic sets which is briefly introduced. For a detailed introduction and discussion, we refer the reader to \cite{BochnakCosteRoy-1998-Springer-RealAlgebraicGeometry} and \cite{BasuPollackRoy-2006-Springer-AlgorithmRealAlgebraicGeometry}. Next, we follow \cite{Brattka-2003-WLCIII-EmporersNewRecursiveness} and define the class of semialgebraic sets as follows.

\begin{definition}
	\label{def:semialgebraic}
	The class of \emph{semialgebraic sets} in $\R^n$ is the smallest class of subsets of $\R^n$ that contains all sets $\{\vec{x}\in\R^n:p(\vec{x})>0\}$ with real polynomials $p:\R^n\rightarrow\R$ and that is further closed under finite intersection and union as well as complement. 
\end{definition}

It follows from the Tarski-Seidenberg Theorem \cite{Tarski-1951-RAND-DecisionMethodElementaryAlgebraGeometry,Seidenberg-1954-AnnMath-NewDecisionMethodElementaryAlgebra} that semialgebraic sets are closed under projection, i.e., if $\sP:\R^n\rightarrow\R^m$ is a projection map and $\sC\subset\R^n$ a semialgebraic set, then $\sP(\sC)\subset\R^m$ is also a semialgebraic set. Further, the interior $\sC^o$, closure $\overline{\sC}$, and therewith also the border $\partial\sC=\overline{\sC}\backslash\sC^o$ of a semialgebraic set $\sC$ are also semialgebraic sets.

It follows that a formula $\Psi$ involving polynomials with logical operations ``$\wedge$'', ``$\vee$'', ``$\neg$'' and quantifiers ``$\forall$'', ``$\exists$'' is a semialgebraic set as well, cf. for example \cite{Cucker-2019-Springer-RecentAdvancesSemialgebraicSets}. Taski found an algorithm that takes a semialgebraic set $\sC$ and a formula $\Psi$ as above as inputs and transforms it into a semialgebraic set without quantifiers in $\R^3$. This allows us to eliminate all quantifiers in semialgebraic sets algorithmically.  Subsequently, the complexity of this elimination process has been reduced significantly compared to the proposed algorithm by Tarski, cf. \cite{Cucker-2019-Springer-RecentAdvancesSemialgebraicSets} and references therein. However, for many applications in information theory it still remains to high.

\section{BSS-Detectability of DoS Attacks}
\label{sec:dos}

In this section we study the detectability of DoS attacks with the help of BSS machines and therewith the capabilities of neuromorphic twins for such tasks.

\subsection{Jammer with Partial Knowledge}
\label{sec:dos_partial}

Here, we show that BSS machines can detect DoS attacks performed by jammers with partial knowledge. For this purpose, we need the following result which is also of interest on its own.
We will construct a simple BSS-computable (linear) bijective function that maps the set $\CHxys$ to a subset in $\R^{|\sX||\sS||\sY|}$. The inverse function will also be BSS-computable. Of particular interest will be image of $\CHxys$ in $\R^{|\sX||\sS||\sY|}$ which is characterized next.

\begin{theorem}
	\label{the:semialgebraic}
	Let $\sX$, $\sY$, and $\sS$ be arbitrary finite alphabets. Then the sets $\Mdos$ and $\Mdos^c$ are semialgebraic sets.
\end{theorem}
\begin{IEEEproof}
	The aim is to construct a bijective function $\vec{t}:\CHxys\rightarrow\R^{|\sX||\sS||\sY|}$ with $\vec{t}$ maps $W$ to $\vec{t}(W)$. For the computability, the same is required for the inverse function.
	
	We map every channel $W\in\CHxys$ to a vector $\vec{t}\in\R^{|\sX||\sS||\sY|}$ as follows:
	\begin{align}
		\label{eq:A1}
		\vec{t} =\! \begin{pmatrix}
			\vec{t}_{1,1,} \\ \vdots \\ \vec{t}_{1,|\sS|} \\ \vec{t}_{2,1} \\ \vdots \\ \vec{t}_{|\sX|,|\sS|}
		\end{pmatrix}
		\;\text{ with }\; 
		\vec{t}_{k,l} =\! \begin{pmatrix}
			t_{k,l}(1) \\ \vdots \\ t_{k,l}(|\sY|)
		\end{pmatrix}
		\!=\! \begin{pmatrix}
			W(1|k,l) \\ \vdots \\ W(|\sY||k,l)
		\end{pmatrix}
	\end{align}
	with $1\leq k\leq|\sX|$ and $1\leq l\leq|\sS|$. Let $\sT_\CH\subset\R^{|\sX||\sS||\sY|}$ be the set of all vectors $\vec{t}$ with
	\begin{equation}
		\label{eq:u1}
		t_{k,l}(r) \geq 0 \qquad 1\leq r \leq|\sY|, 1\leq k \leq|\sX|, 1\leq l \leq|\sS|,
	\end{equation}
	and
	\begin{equation}
		\label{eq:g1}
		\sum_{r=1}^{|\sY|}t_{k,l}(r) = 1 \qquad 1\leq k \leq|\sX|, 1\leq l \leq|\sS|.
	\end{equation}
	This defines a bijective mapping between $\CHxys$ and $\sT_\CH\subset\R^{|\sX||\sS||\sY|}$. The inequalities \eqref{eq:u1} and equations \eqref{eq:g1} are polynomials so that $\sT_\CH\subset\R^{|\sX||\sS||\sY|}$ is a semialgebraic set. 
	
	Similarly, we map every channel $U\in\CHxs$ to a vector $\vec{u}\in\R^{|\sS||\sX|}$. We follow the same line of thinking as above and obtain then a bijective mapping between $\CHxs$ and $\sU_\CH\subset\R^{|\sS||\sX|}$. Likewise, the set $\sU_\CH\subset\R^{|\sX||\sS|}$ is a semialgebraic set. 
	
	Now, we consider the pairs of vectors $(\vec{t},\vec{u})\in\R^{|\sX||\sS||\sY|}\times\R^{|\sS||\sX|}$ and the corresponding set $\sT_\CH\times\sU_\CH$ is then a semialgebraic set in $\R^{|\sX||\sS||\sY|}\times\R^{|\sS||\sX|}$. We define the polynomial $P$ with
	\begin{equation}
		\label{eq:A4}
		\begin{split}
		P(\vec{t},\vec{u}) &= \sum_{k_1,k_2=1}^{|\sX|}\sum_{r=1}^{|\sY|}\Bigg[\sum_{l=1}^{|\sS|}t_{k_1,l}(r)u_{k_2}(l) \\
		&\qquad\qquad\qquad\qquad\qquad -\sum_{l=1}^{|\sS|}t_{k_2,l}(r)u_{k_1}(l)\Bigg]^2
		\end{split}
	\end{equation}
	and consider the set
	\begin{equation*}
		\sN = \Big\{(\vec{t},\vec{u})\in\R^{|\sX||\sS||\sY|}\times\R^{|\sS||\sX|}: P(\vec{t},\vec{u})=0\Big\}.
	\end{equation*}
	Then the intersection
	\begin{equation*}
		\sN\cap\sT_\CH\cap\sU_\CH
	\end{equation*}
	is a semialgebraic set since it is a finite intersection of semialgebraic sets. For a pair of vectors $(\vec{t}_*,\vec{u}_*)$ we have $(\vec{t}_*,\vec{u}_*)\in\sN\cap\sT_\CH\cap\sU_\CH$ if and only if $P(\vec{t}_*,\vec{u}_*)=0$ and $\vec{t}_*\in\sT_\CH$, $\vec{u}_*\in\sU_\CH$. With $W_*$ being the corresponding channel to $\vec{t}_*$ and $U_*$ being the corresponding channel to $\vec{u}_*$, it holds for all $x,x'\in\sX$
	\begin{align*}
		&\sum_{y\in\sY}\bigg[\sum_{s\in\sS}W_*(y|s,x)U_*(s|x')\bigg] \\
		&\qquad\qquad= \sum_{y\in\sY}\bigg[\sum_{s\in\sS}W_*(y|s,x')U_*(s|x)\bigg]
	\end{align*}
	which means that we have $W_*\in\Mdos$.
	
	Reversely, every channel $W\in\Mdos$ with corresponding symmetrizing channel $U$ belongs to the set $\sN\cap\sT_\CH\cap\sU_\CH$ via the mapping to $\vec{t}$ and $\vec{u}$.
	
	We now consider the set
	\begin{equation*}
		\Tdos = \Big\{\vec{t}:\exists\vec{u}\in\R^{|\sS||\sX|}\text{ s.t. }(\vec{t},\vec{u})\in\sN\cap\sT_\CH\cap\sU_\CH\Big\}.
	\end{equation*}
	The algorithm of Tarski, cf. Section \ref{sec:bss_semi}, can now be used to eliminate the existence quantifier algorithmically, i.e., also the set $\Tdos\subset\R^{|\sX||\sS||\sY|}$ is a semialgebraic set. Then there exist polynomials $P_1,...,P_M,Q_1,...,Q_I$ with
	\begin{align*}
		&\Tdos=\Big\{\vec{t}:P_m(\vec{t})=0,1\leq m \leq M \\
		&\qquad\qquad\qquad \text{ and } Q_i(\vec{t})\geq0, 1 \leq i \leq I\Big\}.
	\end{align*}
	Now, we have
	$W\in\Mdos \Leftrightarrow \vec{t}(W)\in\Tdos$
	which shows that $\Mdos$ is a semialgebraic set proving the first part of the theorem.
	
	Now, we move on to the second part. The polynomials $P_1,...,P_M,Q_1,...,Q_I$ from $\Tdos$ can be computed from $\sN\cap\sT_\CH\cap\sU_\CH$ by using the algorithm of Tarski. We define the set
	\begin{equation*}
		\Pos=\Big\{(\vec{t},\vec{u})\in\R^{|\sX||\sS||\sY|}\times\R^{\sS||\sX|}:P(\vec{t},\vec{u})\neq 0\Big\}.
	\end{equation*}
	From the definitions it is clear that this set can actually be expressed as 
	\begin{equation*}
		\Pos=\Big\{(\vec{t},\vec{u})\in\R^{|\sX||\sS||\sY|}\times\R^{\sS||\sX|}:P(\vec{t},\vec{u})> 0\Big\}.
	\end{equation*}
	Obviously, this set is also a semialgebraic set and so is the intersection
	$\Pos\cap\sT_\CH\cap\sU_\CH$.
	We have $(\vec{t},\vec{u})\in\Pos\cap\sT_\CH\cap\sU_\CH$ if and only if for corresponding channels $W\in\CHxys$ and $U\in\CHxs$ it holds 
	\begin{align*}
		&\sum_{x,x'\in\sX}\sum_{y\in\sY}\bigg[\sum_{s\in\sS}W(y|s,x)U(s|x') \\
		&\qquad\qquad\qquad\qquad- \sum_{s\in\sS}W(y|s,x')U(s|x)\bigg]^2 > 0.
	\end{align*}
	This yields the set
	\begin{align*}
		&\Tnodos = \Big\{\vec{t}:\forall\vec{u}\in\R^{|\sS||\sX|} \\
		&\qquad\qquad\qquad\quad\text{ it holds } (\vec{t},\vec{u})\in\Pos\cap\sT_\CH\cap\sU_\CH\Big\}.
	\end{align*}
	From the algorithm of Tarski it follows that all quantifiers can algorithmically be eliminated. As a consequence, $\Tnodos$ is a semialgebraic set and we have $W\in\Mdos^c$ if and only if $\vec{t}\in\Tnodos$. This completes the sketch of proof.
\end{IEEEproof}

\begin{figure}
	\centering
	\scalebox{1}{\includegraphics{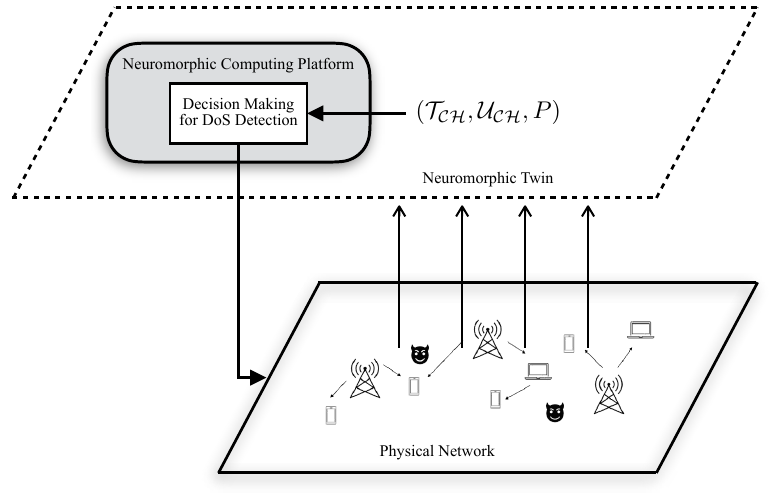}}
	\caption{A neuromorphic twin of the form $(\sT_{\CH},\sU_{\CH},P)$ that provides a suitable description of the physical network which is then used for the detection of DoS attacks. Note that this structure differs to the one of the digital twin.}
	\label{fig:neurotwin}
\end{figure}

\begin{remark}
	In the proof of Theorem \ref{the:semialgebraic}, we provide an ``encoding'' of the physical network into a neuromorphic twin, i.e., a neuromorphic description of the actual communication system. This is provided by \eqref{eq:A1}-\eqref{eq:g1}. The basis for the corresponding computing problem is then given by \eqref{eq:A1}-\eqref{eq:g1} together with \eqref{eq:A4}. The corresponding neuromorphic twin together with the neuromorphic computing hardware is visualized in Fig.~\ref{fig:neurotwin}.
\end{remark}

Now we are in the position to prove the main result which is the detectability of DoS attacks on BSS machines. This shows that the detection task is indeed solvable on neuromorphic computing hardware which provides a positive answer to the question posed above.

\begin{theorem}
	\label{the:decidable}
	Let $\sX$, $\sY$, and $\sS$ be arbitrary finite alphabets. Then there exists a BSS machine $\BSS:\CHxys\rightarrow\{``\text{yes}",``\text{no}"\}$ that outputs $\BSS(\fW)=``\text{yes}"$ if and only if $\fW\in\Mdos$, i.e., the problem of detecting DoS attacks is BSS-decidable.
\end{theorem}
\begin{IEEEproof}[Sketch of Proof]
	A set $\sM$ is a set of all accepted inputs of a BSS machine if and only if it can be expressed as a countable union of semialgebraic sets, cf. \cite{BlumShubSmale-1989-AMS-TheoryComputation,Blum-2004-AMS-ComputingOverReals,BlumCuckerShubSmale-1998-Springer-ComplexityRealComputation}. Note that we want to satisfy a very strong requirement on the detectability: We consider only such BSS machines that always stop for all possible inputs. Such a BSS machine has then two possible outputs: Either the channel (obtained as input) is in the set $\sM$ or not. 
	
	 From Theorem \ref{the:semialgebraic} we know that both $\Mdos$ and $\Mdos^c$ are semialgebraic sets. As a consequence, there exists two BSS machines $\BSS_1$ and $\BSS_2$ with: $\BSS_1$ accepts exactly all channels $\fW$ with $\fW\in\Mdos$ and $\BSS_2$ accepts exactly all channels $\fW$ with $\fW\in\Mdos^c$. 
	
	Now, we can construct a BSS machine $\BSS_*$ as follows: With $\fW\in\CHxys$ as input, we start both BSS machines $\BSS_1$ and $\BSS_2$ in parallel. We know that one of both BSS machines must stop. If $\BSS_1$ stops, then we set $\BSS_*(\fW)=``\text{yes}"$. If $\BSS_2$ stops, then we set $\BSS_*(\fW)=``\text{no}"$. This provides a BSS machine that solves the desired decision task which completes the proof.
\end{IEEEproof}

\subsection{Jammer with Full Knowledge}
\label{sec:dos_full}

Next, we establish the same results for the case of jammers with full knowledge.

\begin{theorem}
	\label{the:semialgebraicfull}
	Let $\sX$, $\sY$, and $\sS$ be arbitrary finite alphabets. Then the sets $\Mdosfull$ and $\Mdosfull^c$ are semialgebraic sets.
\end{theorem}
\begin{IEEEproof}[Sketch of Proof]
	We use the same function $\vec{t}:\CHxys\rightarrow\R^{|\sX||\sS||\sY|}$ as defined in the proof of Theorem \ref{the:semialgebraic}. We further define the function $\vec{u}$ as follows: For $1\leq k \leq |\sX|$, $1\leq l \leq |\sX|$, and $P_{k,l}\in\sP(\sS)$ arbitrary, we define
	\begin{equation*}
		\vec{u} = \begin{pmatrix}
			\vec{u}_{1,1} \\ \vdots \\ \vec{u}_{1,|\sX|} \\ \vec{u}_{2,1} \\ \vdots \\ \vec{u}_{|\sX|,|\sX|}
		\end{pmatrix}\in\R^{|\sS||\sX|^2}
	\end{equation*}
	with
	\begin{equation*}
		\vec{u}_{k,l} = \begin{pmatrix}
			P_{k,l}(1) \\ \vdots \\ P_{k,l}(|\sS|)
		\end{pmatrix}\in\R^{|\sS|}.
	\end{equation*}
	We set
	\begin{align*}
		\overline{\sT}_\CH=\Big\{\vec{u}:\vec{u}_{k,l}(r)\geq0, 1\leq r \leq |\sS|, 1\leq k,l \leq |\sX|\quad \\
		\text{and } \sum_{r=1}^{|\sS|}\vec{u}_{k,l}(r)=1, 1\leq k,l\leq |\sX|\Big\}
	\end{align*}
	and observe that $\overline{\sT}_\CH\subset\R^{|\sS||\sX|^2}$ is a semialgebraic set. We further define the polynomial
	\begin{align*}
		P(\vec{t},\vec{u}) &= \sum_{k_1,k_2=1}^{|\sX|}\sum_{r=1}^{|\sY|}\Bigg[\sum_{l=1}^{|\sS|}\!t_{k_1,l}(r)u_{k_1,k_2}(l) \\
		&\qquad\qquad\qquad\qquad\quad- \sum_{l=1}^{|\sS|}\!t_{k_2,l}(r)u_{k_1,k_2}(l)\Bigg]^{2}
	\end{align*}
	for $(\vec{t},\vec{u})\in\R^{|\sX||\sS||\sY|}\times\R^{|\sS||\sX|^2}$. With this polynomial and the set $\overline{\sT}_\CH$ we can follow the derivation in the proof of Theorem \ref{the:semialgebraic} to show the desired result. The details are omitted for brevity.
\end{IEEEproof}

\begin{theorem}
	\label{the:decidablefull}
	Let $]\sX$, $\sY$, and $\sS$ be arbitrary finite alphabets. Then there exists a BSS machine $\BSS:\CHxys\rightarrow\{``\text{yes}",``\text{no}"\}$ that outputs $\BSS(\fW)=``\text{yes}"$ if and only if $\fW\in\Mdosfull$, i.e., the problem of detecting DoS attacks is BSS-decidable.
\end{theorem}
\begin{IEEEproof}
	The proof is similar to the proof of Theorem \ref{the:decidable} where we now have to use Theorem \ref{the:semialgebraicfull}. The details are omitted for brevity.
\end{IEEEproof}

\section{State and Input Constraints}
\label{sec:constraints}

\begin{figure}
	\centering
	\scalebox{1}{\includegraphics{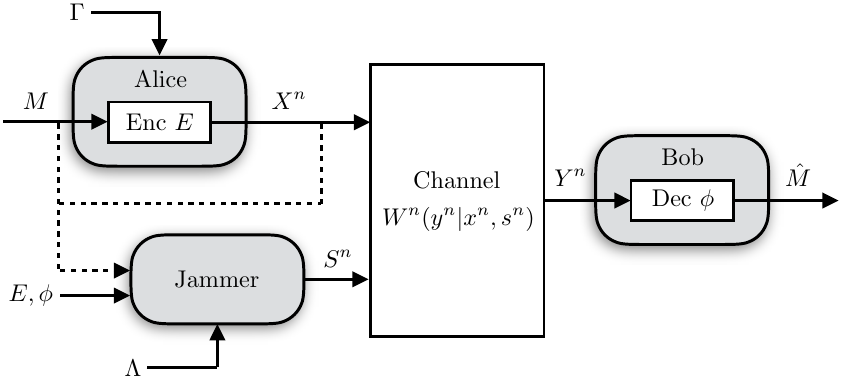}}
	\caption{Communication system with an active adversary subject to state and input constraints. The Jammer is limited to those jamming sequences that satisfy the state constraint $\Lambda$. Similarly, the transmitter can only use such codewords that satisfy the input constraint $\Gamma$.}
	\label{fig:constraints}
\end{figure}

From a practical point of view, it is important to take constraints on the admissible jamming and input sequences into account as visualized in Fig.~\ref{fig:constraints}. A jamming sequence transmitted by an adversary will always be subject to a power constraint. The same applies to the legitimate transmitter, where admissible input sequences are also limited by a corresponding transmit power constraint. 

We will first consider the case where only the jamming sequences are limited by a power constraint. Subsequently, we will generalize this by also taking constraints on the input sequences into account.

\subsection{State Constraints}
\label{sec:constraints_state}

Here, we consider the case of state constraints for the jamming sequences of the adversary. We follow \cite{CsiszarNarayan-1988-AVCConstrainedInputsStates,CsiszarNarayan-1988-AVCRevisited} and introduce the state constraints as follows. For state sequence $s^n=(s_1,...,s_n)$ we define the cost function 
\begin{equation*}
	l(s^n)=\frac{1}{n}\sum_{i=1}^nl(s_i)
\end{equation*}
and assume as in \cite{CsiszarNarayan-1988-AVCConstrainedInputsStates} that $\min_{s\in\sS}l(s)=0$. If we have $\Lambda\geq \max_{s\in\sS}l(s)$ for a given state constraint $\Lambda>0$, then the state constraint is inactive. For what follows, we need the definition of a type. 

\begin{definition}
	\label{def:type}
	The \emph{type} of a sequence $s^n=(s_1,s_2,...,s_n)\in\sS^n$ of length $n$ is a distribution $P\in\sP(\sS)$ defined by $p(a)\coloneqq \frac{1}{n}N(a|s^n)$ for every $a\in\sS$, where $N(a|s^n)$ denotes the number of indices $i$ such that $s_i=a$, $i=1,...,n$. 
\end{definition}

Let $p\in\sP(\sX)$ be fixed but arbitrary and not necessarily a type. Let $\Lambda$ be the power constraint of the jammer. For $W\in\CH(\sX,\sS;\sY)$ let $\sU(W)$ be the set of all matrices that symmetrize the channel (assuming that this set is non-empty).

If for a type $P$, we have 
\begin{equation}
	\label{eq:lambda}
	\Lambda_0(P)=\min_{U\in\sU(W)}\sum_{x\in\sX}\sum_{s\in\sS}P(x)U(s|x)l(s) < \Lambda,
\end{equation}
then the adversary is able to symmetrize the channel as the costs $\Lambda_0$ are less than the state constraint $\Lambda$. As a consequence, there exists no code with positive rate and a DoS attack is possible in this case. We denote the set of channels $W\in\CH(\sX,\sS;\sY)$ such that \eqref{eq:lambda} is satisfied by $\Msyms$, where this set is defined for general $P\in\sP(\sX)$ and not necessarily types only. Similarly, $\Msymse$ is the corresponding set of channels where \eqref{eq:lambda} is satisfied with $\leq\Lambda$.

On the other hand, if for a type $P$, we have
\begin{equation}
	\label{eq:lambda2}
	\Lambda_0(P)=\min_{U\in\sU(W)}\sum_{x\in\sX}\sum_{s\in\sS}P(x)U(s|x)l(s) > \Lambda,
\end{equation}
then a code  with positive rate exists, cf. \cite{CsiszarNarayan-1988-AVCRevisited}, and a DoS attack is not possible. Intuitively, this is the case because the costs $\Lambda_0$ for symmetrizing the channel would exceed the available cost budget $\Lambda$. 

Note that the remaining case of equality, i.e., for $\Lambda_0(P)=\Lambda$, remains open. We have the following result.

\begin{theorem}
	\label{the:states_semialgebraic}
	Let $\sX$, $\sY$, and $\sS$ be arbitrary finite alphabets. Further, let $P\in\sP(\sX)$ and $\Lambda>0$ be arbitrary but fixed. Then the sets $\Msyms$ and $\Msymse$ are semialgebraic sets.
\end{theorem}
\begin{IEEEproof}
	It is sufficient to show the result for $\Msyms$. The other case $\Msymse$ can be proven in the same way. The proof heavily relies on the proof of Theorem~\ref{the:semialgebraic} and we will, accordingly, point out the modifications. 
	
	We use the same vector $\vec{t}\in\R^{|\sX||\sS||\sY|}$ as in Theorem \ref{the:semialgebraic}, which defines the set $\sT_\CH(P,\Lambda)$. Similarly, also use $\vec{u}\in\R^{|\sS||\sX|}$ and define one new inequality for $\vec{u}$ to account for the state constraints as follows:
	\begin{equation}
		\label{eq:state_uneu}
		\sum_{k=1}^{|\sX|}\sum_{r=1}^{|\sS|}P(k)u_k(r)l(r) < \Lambda.
	\end{equation}
	This inequality \eqref{eq:state_uneu} is linear in $\vec{u}$ and, therewith, defines a semialgebraic set. With this, the rest of the proof follows as in Theorem \ref{the:semialgebraic} and is omitted for brevity.
\end{IEEEproof}
\addspace

With this, we immediately obtain the following result. 

\begin{theorem}
	\label{the:states_bss}
	Let $\sX$, $\sY$, and $\sS$ be arbitrary finite alphabets. Further, let $P\in\sP(\sX)$ and $\Lambda>0$ be arbitrary but fixed. Then the sets $\Msyms$ and $\Msymse$ are both BSS-decidable. 
\end{theorem}
\begin{IEEEproof}
	With Theorem \ref{the:states_semialgebraic}, the proof follows as in Theorem~\ref{the:decidable} and is omitted for brevity.
\end{IEEEproof}
\addspace

Some discussion is in order. 
\begin{enumerate}
	\item These results do not provide a complete characterization of DoS attacks, since it is not clear, what happens in the case of equality in \eqref{eq:lambda}.
	\item If  $P\in\sP(\sX)$ is a type and fixed, condition \eqref{eq:lambda} is sufficient for a DoS attack being possible. Note that the code needs to have the property that all codewords must be of type $P$. Note that this is only sufficient but not necessary, since it could be the case that there is another type $\hat{P}\in\sP(\sX)$ for which a DoS is not possible.
	\item Condition \eqref{eq:lambda2} is sufficient for a DoS attack not being possible if $P\in\sP(\sX)$ is a type and fixed. This corresponds to the complementary set $\Msymge$ and, accordingly, this set is also semialgebraic and therewith BSS-decidable. This means we immediately obtain the following result.
\end{enumerate}

\begin{theorem}
	\label{the:states_bss2}
	Let $\sX$, $\sY$, and $\sS$ be arbitrary finite alphabets. Further, let $P\in\sP(\sX)$ and $\Lambda>0$ be arbitrary but fixed. Then the set $\Msymge$ is BSS-decidable. 
\end{theorem}

\subsection{Input and State Constraints}
\label{sec:constraints_input}

Finally, we address the most general case by further considering a constraint on the input sequences. For $x^n=(x_1,...,x_n)$ we define the cost function 
\begin{equation*}
	g(x^n)=\frac{1}{n}\sum_{i=1}^ng(x_i)
\end{equation*}
and assume as in \cite{CsiszarNarayan-1988-AVCConstrainedInputsStates} that $\min_{x\in\sX}g(x)=0$. If we have $\Gamma\geq \max_{x\in\sX}g(x)$ for a given input constraint $\Gamma>0$, then the input constraint is inactive. We further define
\begin{equation*}
	g(P) = \sum_{x\in\sX}P(x)g(x).
\end{equation*}

Based on the discussion above for the case of state constraints only, we can now state conditions for the general case including both state and input constraints. We define 
\begin{equation*}
	\Mdenial(\Gamma,\Lambda) = \big\{W: \max_{P\in\sP(\sX),g(P)\leq\Gamma}\Lambda(P,W)<\Lambda\big\}
\end{equation*}
with 
\begin{equation*}
	\Lambda(P,W)=\min_{U\in\sU(W)}\sum_{x\in\sX}\sum_{s\in\sS}P(x)U(s|x)l(s),
\end{equation*}
cf. \eqref{eq:lambda} and \eqref{eq:lambda2}, which describes the set of all channels for which a DoS attack is possible under state constraint $\Lambda$ and input constraint $\Gamma$ for a Jammer with partial knowledge. Similarly, we can define the corresponding set $\Mdenialfull(\Gamma,\Lambda)$ for a Jammer with full knowledge. We have the following result.

\begin{theorem}
	\label{the:input}
	Let $\sX$, $\sY$, and $\sS$ be arbitrary finite alphabets. Further, let $\Lambda>0$ and $\Gamma>0$ be arbitrary but fixed. Then the sets $\Mdenial(\Gamma,\Lambda)$ and $\Mdenialfull(\Gamma,\Lambda)$ are both BSS-decidable. 
\end{theorem}
\begin{IEEEproof}
	We can prove the desired result by extending the proof for the case of only state constraints in Theorem \ref{the:states_semialgebraic}. 
	
	In addition to the vectors $\vec{t}\in\R^{|\sX||\sS||\sY|}$ and $\vec{u}\in\R^{|\sS||\sX|}$, we consider the vector 
	\begin{equation*}
		\vec{p} = \begin{pmatrix}
			p(1) \\ \vdots \\ p(|\sX|)
		\end{pmatrix}\in\R^{|\sX|}, \quad p\in\sP(\sX)
	\end{equation*}
	with
	\begin{align*}
		&p(l) \geq 0, \quad 1\leq l \leq |\sX|, \\
		&\sum_{l=1}^{|\sX|}p(l)=1, \text{ and}\\
		&\sum_{l=1}^{|\sX|}p(l)g(l) \leq \Gamma.
	\end{align*}
	We observe that this defines a semialgebraic set  in $\R^{|\sX|}$ which we denote as $\sP(g,\Gamma)$. Now, the variables
	\begin{equation*}
		(\vec{t},\vec{u},\vec{p})\in\R^{|\sX||\sS||\sY|}\times\R^{|\sS||\sX|}\times\R^{|\sX|}
	\end{equation*}
	with $(\vec{t},\vec{u})\in\sT_\CH(P,l,\Lambda)\times\sU_\CH$ and $\vec{p}\in\sP(g,\Gamma)\subset\R^{|\sX|}$. Now, the set of vectors $(\vec{t},\vec{u},\vec{p})$ with these properties are a semialgebraic set which we denote as $\sA(g,\Gamma,l,\Lambda)$. 
	
	We observe  that $W\in\Mdenial(\Gamma,\Lambda)$ is true if and only if for all $\vec{p}\in\R^{|\sX|}$ there exists a $\vec{u}\in\R^{|\sS||\sX|}$ with $\vec{p}\in\sP(g,\Gamma)$ and $\vec{t}\in\sT_\CH(P,l,\Lambda)$ so that $(\vec{t},\vec{u},\vec{p})$ is in $\sA(g,\Gamma,l,\Lambda)$. Since $\sA(g,\Gamma,l,\Lambda)$ is a semialgebraic set, the set of all these $\vec{t}$ is a semialgebraic set and therewith also the set $\Mdenial(\Gamma,\Lambda)$ is a semialgebraic set. This allows us to use the Tarski-Seidenberg-Theorem to eliminate all the quantifiers in $\Mdenial(\Gamma,\Lambda)$.
	
	The derivation for $\Mdenialfull(\Gamma,\Lambda)$ follows accordingly which completes the proof.
\end{IEEEproof}

\section{Discussion}
\label{sec:discussion}

In this paper, we have studied the detectability of DoS attacks. In \cite{BocheSchaeferPoor-2020-TSP-DoS} it has been shown that this problem cannot be solved algorithmically on Turing machines. It has subsequently been shown in \cite{BocheSchaeferPoor-2021-ICC-AlgorithmicDetectionAdversarialAttack,BocheSchaeferPoor-2021-TNET-AlgorithmicSolvability} that Turing machines are not capable of algorithmically detecting DoS attacks on communication systems with feedback. These results show the limitations of today's information processing hardware and the framework of digital twins. Even if the best possible digital twins and the best possible digital computing platform are used, the detectability of DoS attacks remains impossible.

Recently, the limitations of Turing machines and therewith of digital twins have been observed for communication systems. In particular, it has been shown that the deterministic capacity of arbitrarily varying channel \cite{BocheSchaeferPoor-2020-TIFS-EffectivePerformanceEvaluation}, the capacity of finite state channels \cite{ElkoussPerezGarcia-2018-Nature-Uncomputable,BocheSchaeferPoor-2020-CIS-NonComputabilityFSC}, and the identification capacity with feedback \cite{BocheSchaeferPoor-2020-TIT-IDF} are not computable on Turing machines. To this end, it has further been shown that optimal information processing schemes for several problems are not algorithmically constructible. This shows the fundamental limits of a computer-aided information processing design. Such a behavior has been shown for compound and block fading channels in \cite{BocheSchaeferPoor-2020-TSP-CommunicationChannelUncertainty} and \cite{BocheSchaeferPoor-2021-ICASSP-CommunicationBlockFading}. In addition to that, there is no Turing machine that takes a discrete memoryless channel (DMC) as an input and then algorithmically computes optimal, i.e., capacity-achieving, codes \cite{BocheSchaeferPoor-2021-ICC-AlgorithmicConstructabilityCodes}. This also motivates the analysis of more general hardware platforms, i.e., those that go beyond the standard digital hardware, from a computing perspective; even if such hardware platforms cannot be built with today's technology. 

BSS machines provide a suitable computing model for neuromorphic processors and therewith provide the basis for corresponding neuromorphic twins. Recently, there has been significant progress in the development in such neuromorphic processors. In particular, prototypes have been developed with an analog signal processing that show a significantly reduced energy consumption compared to digital processors. This already makes them particularly interesting for machine learning approaches in signal processing as solutions on digital hardware have shown a huge energy consumption. Further, BSS machines have been proposed to be a suitable computing model for biocomputing in living organisms on all scales ranging from computing in cells to computing in the brain \cite{Grozinger-2019-Biocomputing}. In addition to this, in this paper we observe additional new features of neuromorphic computing solutions: They are capable to solve practically relevant problems with guaranteed performance that are impossible to solve on traditional digital hardware. 

In this paper, we have considered BSS machines and have shown that DoS attacks become indeed detectable on such computing models that are able to process and store arbitrary real numbers. The results reveal that the symmetrizability condition can be expressed as independent polynomial equations and inequalities so that the question of detectability is then completely characterized by polynomials and therewith by semialgebraic sets. The set of AVCs that satisfy this condition further involves an existence quantifier which can be eliminated by the Tarski algorithm. Thus, there exists a BSS machine that can decide for every AVC whether or not a DoS attack is possible. In contrast to this, such a elimination of quantifiers is not possible on Turing machines provides a reason that this decision problem is not solvable on Turing machines.

The problem of detection DoS attacks on communication systems is relevant by itself but is further of fundamental importance for the verification of trustworthiness. This gives evidence that also for trustworthiness verification, the underlying hardware platform must be taken into account for the implementation. For example, this means that even if it has been shown that a certain procedure  is trustworthy, it can become non-trustworthy if it need to be implemented on a digital hardware where such a corresponding implementation can never be trustworthy \cite{BocheSchaeferPoorFettweis-2022-ICC-Trustworthiness}. This is exactly the behavior that we have seen for the detection of DoS attacks. This makes the use of neuromorphic processors very interesting for the implementation of trustworthiness.

\balance



\end{document}